\documentclass[12pt]{article}
\usepackage{epsfig,fullpage}
\newcommand{\lapprox}{\raisebox{-0.5ex}{$\
\stackrel{\textstyle<}{\textstyle\sim}\ $}}
\newcommand{\gapprox}{\raisebox{-0.5ex}{$\
\stackrel{\textstyle>}{\textstyle\sim}\ $}}
\newcommand{\One}{1\kern-4.5pt1}
\begin{document}
\begin{center}

\begin{flushright}
     SWAT/04/395 \\
DUKE-TH-04-261\\
April 2004
\end{flushright}
\par \vskip 10mm

\vskip 1.2in

\begin{center}
{\LARGE\bf
Non-Compact QED$_3$ with $N_f=1$ and $N_f=4$} 

\vskip 0.7in
S.J. Hands $\!^{a,b}$, J.B. Kogut $\! ^c$, L. Scorzato$\! ^d$  and
 C.G. Strouthos $\!^{e,f}$ \\
\vskip 0.2in

$^a\,${\it Department of Physics, University of Wales Swansea,\\
Singleton Park, Swansea, SA2 8PP, U.K.} \\
$^b\,${\it Institute for Nuclear Theory, University of Washington,\\
Box 351550, Seattle WA 98195, U.S.A.}\\
$^c\,${\it Department of Physics, University of Illinois at Urbana-Champaign,\\
Urbana, IL 61801-3080, U.S.A.}\\
$^d\,${\it Institut f\"ur Physik, Humboldt Universit\"at zu Berlin, 12489
Berlin, Germany,}\\
$^e\,${\it Department of Physics, Duke University, Durham, NC 27708,
U.S.A.}\\
$^f\,${\it Division of Science and Engineering, Frederick Institute of
Technology, \\Nicosia 1303, Cyprus.}
\end{center}

\vskip 1.0in 
{\large\bf Abstract}
\end{center}
\noindent
We present numerical results for 
non-compact three-dimensional QED for numbers of flavors $N_f=1$ and $N_f=4$.
In particular, we address the issue of whether chiral symmetry is spontaneously
broken in the continuum limit, and obtain a positive answer for $N_f=1$, with a
dimensionless condensate estimated to be
$\beta^2\langle\bar\psi\psi\rangle\simeq O(10^{-3})$, implying that the critical
number of flavors $N_{fc}>1$. We also compare
the $N_f=1$ and $N_f=4$ models by analysing the transition from strong to weak
coupling behaviour using an equation of state based on a continuous phase
transition. While some qualitative differences emerge, it appears difficult 
to determine whether $N_f=4$ lies above or below $N_{fc}$.

\newpage

\section{Introduction and Motivation}

The nature of the ground state of Quantum Electrodynamics in 2+1 dimensions
and its dependence on the number of fermion species $N_f$ has been an important
problem for non-perturbative field theorists for 20 years \cite{pisarski}.
The question is the value of critical number of flavors $N_{fc}$ such that 
for $N_f<N_{fc}$ the theory's chiral symmetry is spontaneously broken,
resulting in a dynamically generated mass for the fermion.
Over the years the main method of attack has been to solve the Schwinger-Dyson 
equation for the fermion self-energy $\Sigma(k)$, to see if
$\lim_{k\to0}\Sigma(k)\not=0$; this has resulted in a variety of predictions for
$N_{fc}$ ranging between $32/\pi^2$ and $\infty$, with most recent attempts 
yielding $N_{fc}\simeq4$ \cite{SD}. Over the same period there have also been 
attempts to decide the issue by numerical simulation of non-compact lattice QED
\cite{DKK,lattice,HK}. Numerical approaches are hindered by large finite volume
effects, and the intrinsic smallness of the relevant signal, the so-called
dimensionless condensate $\beta^2\langle\bar\psi\psi\rangle$ (to be defined in
what follows) in the continuum limit: 
in a recent study \cite{HKS} 
of the model with $N_f=2$ we placed an upper bound on this
quantity of $O(10^{-4})$, but even this does not exclude the predictions of some
self-consistent approaches with $N_{fc}>2$ \cite{AW,Pieter}.

Interest in the problem
has recently been revived by the suggestion that QED$_3$ may be 
an effective theory for the underdoped and non-superconducting region of 
the phase diagram of high-$T_c$ superconducting cuprate compounds
\cite{Tesanovic, Herbut}.
In brief, superconductivity in these substances is confined to planes defined by 
CuO$_2$ layers, thus motivating a 2+1$d$ description. 
The superconducting order parameter has a $d$-wave symmetry, implying that there
are four nodes in the gap function $\Delta(\vec k)$ as the Fermi surface (which
in 2+1$d$ is a curve) is circumnavigated. At each node the low-energy
quasiparticle excitations obey an approximately linear dispersion relation
$E(\vec k)\propto \vert\vec k-\vec k_{node}\vert$, 
with the result that it is possible to rewrite the action
for eight distinct low energy species (spin up and spin down at each of
four nodes) in a relativistically invariant form in terms of $N_f=2$ species
of four-component Dirac spinor\footnote{For phenomenologically relevant models
the action for each individual flavor exhibits a spatial anisotropy, a feature 
ignored in this paper.}. The next link in the chain is the observation 
that disruption of superconducting order in a planar system occurs via
condensation of vortex singularities in the phase $\phi$ of the
U(1)-valued order parameter, while
$\vert\Delta\vert$ remains unchanged. 
Because the order parameter field is doubly charged since it arises from
electron pairing, in the presence of a distribution of 
vortices it is impossible in general 
to reabsorb $\phi$ into the definition of the quasiparticle fields
while leaving their wavefunction single-valued. There is, however,
a particular gauge \cite{FT} in which the problem can be recast in
terms of single-valued quasiparticle fields interacting with a real-valued 
vector field.
The component of the vector field that is minimally coupled to the
quasiparticles remains massless under quantum corrections (vacuum
polarisation due to virtual particle hole pairs), and is identified with 
the photon of QED$_3$\footnote{In fact, the photon can be regarded as the
Goldstone boson associated with vortex condensation \cite{Alex}.}. 
It is then argued \cite{Tesanovic,Herbut} 
that the effective action
for the photon fluctuations is of the non-compact form $F_{\mu\nu}^2$, with an 
effective ``electric charge'' coupling photons to quasiparticles proportional to
the vortex disorder parameter, which is 
non-zero as soon as vortices unbind, ie. {\em outside} the superconducting
region of the phase diagram.

If QED$_3$ is a relevant effective theory for cuprates, then
the abstract theoretical problem of
the value of $N_{fc}$ assumes concrete phenomenological reference. If
$N_{fc}>2$, then the theory is chirally broken at zero temperature. On
retranslating from the Dirac spinor basis to the original electron degrees of
freedom, the chiral order parameter is reinterpreted as an order parameter for
spin density waves, whose wavevector gets shorter and shorter as doping is
decreased, until at zero doping the N\'eel antiferromagnetic state is recovered
\cite{Herbut}. This picture therefore predicts the existence of a phase boundary
between superconducting (dSC) and antiferromagnetic (AFM) phases at some non-zero doping in
the zero temperature limit. If, on the other hand, $N_{fc}<2$, the chirally
symmetric ground state manifests itself as a tongue of ``pseudogap'' phase
separating dSC from AFM, in which normal Fermi liquid properties may be modified
as a result of a non-perturbative anomalous dimension for the fermion field 
\cite{Tesanovic}.

In order to motivate further
study, let us review two arguments concerning $N_{fc}$
which are independent of Schwinger-Dyson analyses.
A recent discussion of similar issues can be found in \cite{AW}.
The first due, to Appelquist, Cohen and Schmaltz \cite{appelquist}, postulates
an inequality 
\begin{equation}
f_{IR}\leq f_{UV}
\label{eq:cbound}
\end{equation}
between thermodynamic free energy densities (ie. pressures)  due to the
weakly-interacting degrees of freedom characteristic of respectively
long and short distance regimes. Since $f$ is
simply related to the number of weakly interacting degrees of freedom, this can
be easily checked for QED$_3$. In the UV limit $f_{UV}\propto1+{3\over4}(4N_f)$,
where 1 counts the photon, which has only one physical polarisation in 2+1$d$, 
and the second term, which includes a numerical
factor arising from Fermi-Dirac statistics, counts free fermions. In the IR
limit for $N_f<N_{fc}$ there is still a photon, but 
chiral symmetry breaking $\mbox{U}(2N_f)\to\mbox{U}(N_f)\otimes\mbox{U}(N_f)$
implies $2N_f^2$ Goldstone modes, so $f_{IR}\propto1+2N_f^2$. Eqn. (\ref{eq:cbound})
then yields the bound $N_{fc}\leq{3\over2}$. It should be pointed out, however,
that this argument may not be applicable to a 2+1$d$ system, since in this 
low dimensionality bosonic modes exhibit strong fluctuations and hence
are never weakly interacting \cite{Mavro}.

Another argument starts from the similarity between QED$_3$ and the
2+1 dimensional Thirring model defined by the Lagrangian
\begin{equation}
{\cal L}_{Thir}=\bar\psi_i(\partial{\!\!\!/\,}+m)\psi_i+
{G^2\over{2N_f}}(\bar\psi_i\gamma_\mu\psi_i)^2.
\end{equation}
In the limit $G^2\to\infty$ there is a massless vector meson, ie. a
$\psi\bar\psi$ bound state
with the same quantum numbers as the photon; moreover
both models have identical quantum corrections when calculated in an expansion
in powers of $N_f^{-1}$ \cite{Espriu}. However, the resemblance may also be 
non-perturbative in $N_f^{-1}$; Itoh {\em et al\/} \cite{Itoh} pointed out that 
in a certain non-local gauge the self-consistent gap equation for chiral
symmetry breaking in the Thirring model coincides in the limit $G^2\to\infty$
with that of QED$_3$, implying a vanishing chiral order parameter for some 
critical
$N_{fc}^\prime$. In the Thirring case, however, the vanishing of the gap also
happens
for $N_f<N_{fc}^\prime$, $G^2=G_c^2<\infty$, and signals a UV fixed point of the
renormalisation group at which a non-trivial continuum limit may be taken.
This picture has received support from numerical simulations with finite $N_f$, 
which suggest that $4<N_{fc}^\prime<6$ \cite{Thirring1,Thirring2,Thirring3}.
Since the pattern of global symmetry breaking is the same for both models,
it is tempting to postulate that the IR fixed point
behaviour of QED$_3$ coincides with the UV behaviour of Thirring in the
$G^2\to\infty$ limit, and hence $N_{fc}=N_{fc}^\prime\sim5$.

Since neither argument is obviously nonsense, and since they yield predictions
for $N_{fc}$ which differ significantly
both from each other and from the Schwinger-Dyson 
approach, it is clear that consensus on this issue has yet to emerge, and that
further numerical efforts are justified.
In \cite{HKS} we were unable to find any signal for chiral symmetry breaking,
but also unfortunately were not able to definitively exclude it being present
with the tiny value predicted in the self-consistent approach \cite{AW,Pieter}.
In this paper we build on that work by performing extensive
simulations of non-compact QED$_3$ with $N_f=1$ and $N_f=4$, the hope being
that if these two
values were to straddle $N_{fc}$, then some difference in their behaviour,
either quantitative or qualitative, may
show up. The lattice formulation and simulation 
method is reviewed in the next section, and the numerical results presented
in Sec.~\ref{sec:results}. We shall see that for $N_f=1$
there is convincing evidence for
broken chiral symmetry, implying $N_{fc}>1$; unfortunately, while we can uncover 
some qualitative differences between $N_f=1$ and $N_f=4$, there is no decisive
evidence that they have different ground states in the continuum limit. 

\section{The Model and the Simulation}

The action of the lattice model we study is 
\begin{eqnarray}
S &=&\frac{\beta}{2} \sum_{x,\mu<\nu} \Theta_{\mu \nu}(x) \Theta_{\mu \nu}(x)
+ \sum_{i=1}^N \sum_{x,x^\prime} {\bar \chi}_i(x) M(x,x^\prime)
\chi_i(x^\prime)
\label{eq:action}\\
\Theta_{\mu \nu}(x) &\equiv& \theta_{x\mu}+\theta_{x+\hat\mu,\nu}
-\theta_{x+\hat\nu,\mu}-\theta_{x\nu}\nonumber\\
M(x,x^\prime) &\equiv&
m_0\delta_{x,x^\prime}+\frac{1}{2} \sum_\mu\eta_{\mu}(x)
[\delta_{x^\prime,x+\hat \mu} U_{x\mu}
-\delta_{x^\prime,x-\hat \mu} U_{x-\hat \mu,\mu}^\dagger].\nonumber
\end{eqnarray}
This describes interactions between $N$ flavors of
Grassmann-valued staggered fermion fields
$\chi,\bar\chi$ defined on the sites $x$ of a three-dimensional cubic lattice, and
real photon fields $\theta_{x\mu}$ defined on the link between nearest neighbour
sites $x$, $x+\hat\mu$. Since $\Theta^2$ is unbounded from above, (\ref{eq:action})
defines a
non-compact formulation of QED; note however that to ensure local gauge
invariance the fermion-photon interaction is encoded via the compact connection
$U_{x\mu}\equiv\exp(i\theta_{x\mu})$, with $U_{x+\hat\mu,-\mu}=U^*_{x\mu}$. 
In the fermion kinetic matrix $M$ the Kawamoto-Smit phases
\begin{equation}
\eta_\mu(x)=(-1)^{x_1+\cdots+x_{\mu-1}}
\label{eq:KS}
\end{equation}
are designed to
ensure relativistic covariance in the continuum limit, and $m_0$ is the bare
fermion mass.

If the physical lattice spacing is denoted $a$, then
in the continuum limit $a\partial\to0$, (\ref{eq:action}) can be shown to be
equivalent up to terms of $O(a^2)$ to
\begin{equation}
S=\sum_{j=1}^{N_f}\bar\psi^j[\gamma_\mu(\partial_\mu+igA_\mu)+m]\psi^j
+{1\over4}F_{\mu\nu}F_{\mu\nu}
\label{eq:Scont}
\end{equation}
ie. to continuum QED in 2+1 euclidean dimensions,
with $\psi,\bar\psi$ describing $N_f$ flavors of 4 component Dirac spinor acted
on by 4$\times$4 matrices $\gamma_\mu$, and
$N_f\equiv2N$. The continuum photon field is related to the lattice field via 
$\theta_{x\mu}=agA_\mu(x)$, with dimensional coupling strength $g$ given 
by $g^2=(a\beta)^{-1}$, and the field strength
$F_{\mu\nu}=\partial_\mu A_\nu-\partial_\mu A_\nu$. The continuum limit is thus
taken when the dimensionless inverse coupling $\beta\to\infty$.

As reviewed in \cite{HKS}, for $a>0$ in the chiral limit
the lattice action (\ref{eq:action}) retains only a remnant of the 
U(2$N_f$) global symmetry of (\ref{eq:Scont})
under global chiral/flavor rotations, namely a
U($N)\otimes\mbox{U}(N)$ symmetry which is broken to U($N$) either explicitly by
the bare mass $m\not=0$, or spontaneously by a chiral condensate
$\langle\bar\chi\chi\rangle\not=0$, in which case the spectrum contains
$N^2$ exact Goldstone modes. It is expected that the symmetry breaking
pattern  $\mbox{U}(2N_f)\to\mbox{U}(N_f)\otimes\mbox{U}(N_f)$ is restored
in the continuum limit, implying the existence of an additional 7$N^2$
approximate Goldstone modes whose masses vanish as $\beta\to\infty$.

The simulations in this paper were performed using two different algorithms. 
The model with $N_f=4$ is studied by exact simulations of the action
(\ref{eq:action}) with $N=2$ using the hybrid Monte Carlo algorithm, exactly as 
done for $N_f=2$ in \cite{HKS}. The model with $N_f=1$ is simulated using 
$\mbox{det}^{1\over2}M$ as a functional measure using a 
hybrid molecular dynamics R algorithm \cite{Gottlieb}, which generates a Markov
chain of representative configurations by evolution of a system of stochastic
differential equations with timestep $\Delta\tau$. The algorithm has a
systematic error $\propto N^2\Delta\tau^2$; in our work we use values 
of $\Delta\tau$ ranging from 0.1 down to 0.0025, the value required getting
smaller  as $\beta$ decreases, $L_s$ increases, and $m_0$ decreases,
and have
checked that this is sufficient to render systematic errors smaller than
statistical ones. 
Beyond this technical point, however, we should point out the
conceptual issue that simulations with fractional $N$ do not reproduce the
physics of a local fermion bilinear action except in the deep continuum limit
when flavor symmetries are restored. It remains an open issue within the lattice
QCD community whether this is a source of significant systematic error at
achievable lattice spacings
\cite{Hasenfratz}. Whilst we make no attempt to address this issue in the
current study, we should not exclude the possibility that delayed flavor
symmetry restoration could lead to a misleading prediction for the ground state
properties at finite $\beta$ \cite{HKS}.  In Sec.~\ref{sec:pions} below we shall
present evidence for flavor symmetry violating effects of just $O(10\%)$ at
the weakest couplings examined.

To check if chiral symmetry is broken in continuum QED$_3$, we need to 
establish whether, on a system of finite extent $L=L_s\times a$
\begin{equation}
\lim_{\beta\to\infty}\lim_{\beta m_0\to0}\lim_{L_s/\beta\to\infty}
\beta^2\langle\bar\chi\chi\rangle\not=0.
\label{eq:limits}
\end{equation}
The order of limits is important: since a continuous symmetry never breaks
spontaneously on a finite volume we must always work with $m\not=0$, and only
attempt to take the chiral limit once the thermodynamic limit is reached. Since
for an asymptotically-free theory like QED$_3$ the UV behaviour is governed by the
gaussian fixed point at the orgin, all physical quantities are expressible in
terms of the scale set by $g$, and hence to compare data taken at different
lattice spacings, it is natural to use the combinations
$\beta m_0$, $L_s/\beta$ and $\beta^2\langle\bar\chi\chi\rangle$ corresponding
to the dimensionless combinations $m/g^2$, $Lg^2$, and
$g^{-4}\langle\bar\psi\psi\rangle$. As the continuum limit is approached data 
taken at different $\beta$ should collapse onto a universal curve when
plotted in these units.

Taking all three limits of equation (\ref{eq:limits}) is extremely demanding
computationally, as revealed in \cite{HKS}, where in practise we were only able
to place an upper bound on the condensate for $N_f=2$ of
$\beta^2\langle\bar\chi\chi\rangle\leq5\times10^{-5}$. The small size of
the dimensionless condensate in this case implies that a large separation of
scales $\xi\gg a$ may be taking place, and therefore a much larger infra-red
cutoff $L\gg\xi$
required to probe chiral symmetry breaking if it is indeed present
\cite{GusyninReenders}. 
The small number is not, however, inconsistent with current estimates based on
solutions of the Schwinger-Dyson equations \cite{AW,Pieter}. It therefore
appears unlikely that the critical value $N_{fc}$ can be established using
current computational resources if $N_{fc}>2$.

In this paper our aims are twofold. First, since the self-consistent solutions
suggest an exponential suppression $\langle\bar\psi\psi\rangle\sim\exp(-f(N_f))$,
where $f$ is an increasing function of $N_f$ for $N_f<N_{fc}$, the dimensionless
condensate for $N_f=1$ may well be much larger and hence easier to measure using
current resources. Secondly, we will look for qualitative differences between
data taken with $N_f=1$ and $N_f=4$ to see what evidence if any
can be found that they lie on
opposite sides of $N_{fc}$, by fitting the condensate data to a global equation
of state based on a chiral symmetry restoring transition at finite
$\beta_c(N_f)$. For $N_f<N_{fc}$ this ``critical'' coupling  need not correspond to a true phase
transition, but instead could mark a crossover from strong to weak-coupling
behaviour.

\section{Simulation Results}
\label{sec:results}

For $N_f=1$ we have results from lattice volumes $24^3$ 
($0.25\leq\beta\leq0.45$),
$32^3$ ($0.45\leq\beta\leq0.65$), $44^3$ ($\beta=0.55$), $48^3$ ($\beta=0.65$), and
in addition five studies at $\beta=0.90$ 
on $16^3$, $24^3$, $36^3$, $54^3$ and $80^3$. For $N_f=4$ we
have $16^3$ ($0.15\leq\beta\leq0.2$), $24^3$ ($0.15\leq\beta\leq0.225$), $32^3$
($0.4\leq\beta\leq0.6$), and $48^3$ ($\beta=0.4$). 
For the most part we examined 
mass values in
the range $0.003\leq m_0\leq0.025$. For each parameter set we took typically
200 - 300 trajectories of mean length $\bar\tau=1.0$.
Several different computing facilities were used, so an estimate of the cpu
resource required is approximate at best, but a figure of 12 months running 
on a cluster of 30 1.2 GHz processors is not unreasonable.
In Figs.~\ref{fig:gplot1} ($N_f=1$) and \ref{fig:gplot4} ($N_f=4$)
we show the datasets thus
obtained for $\langle\bar\chi\chi\rangle$ as a function of $\beta$ and $m_0$.
The lines are fits to a global equation of state to be described in
Sec.~\ref{sec:global}.

\subsection{Chiral Condensate in Chiral and Continuum Limits}

\begin{figure}[p]

                \centerline{ \epsfysize=4.2in
                             \epsfbox{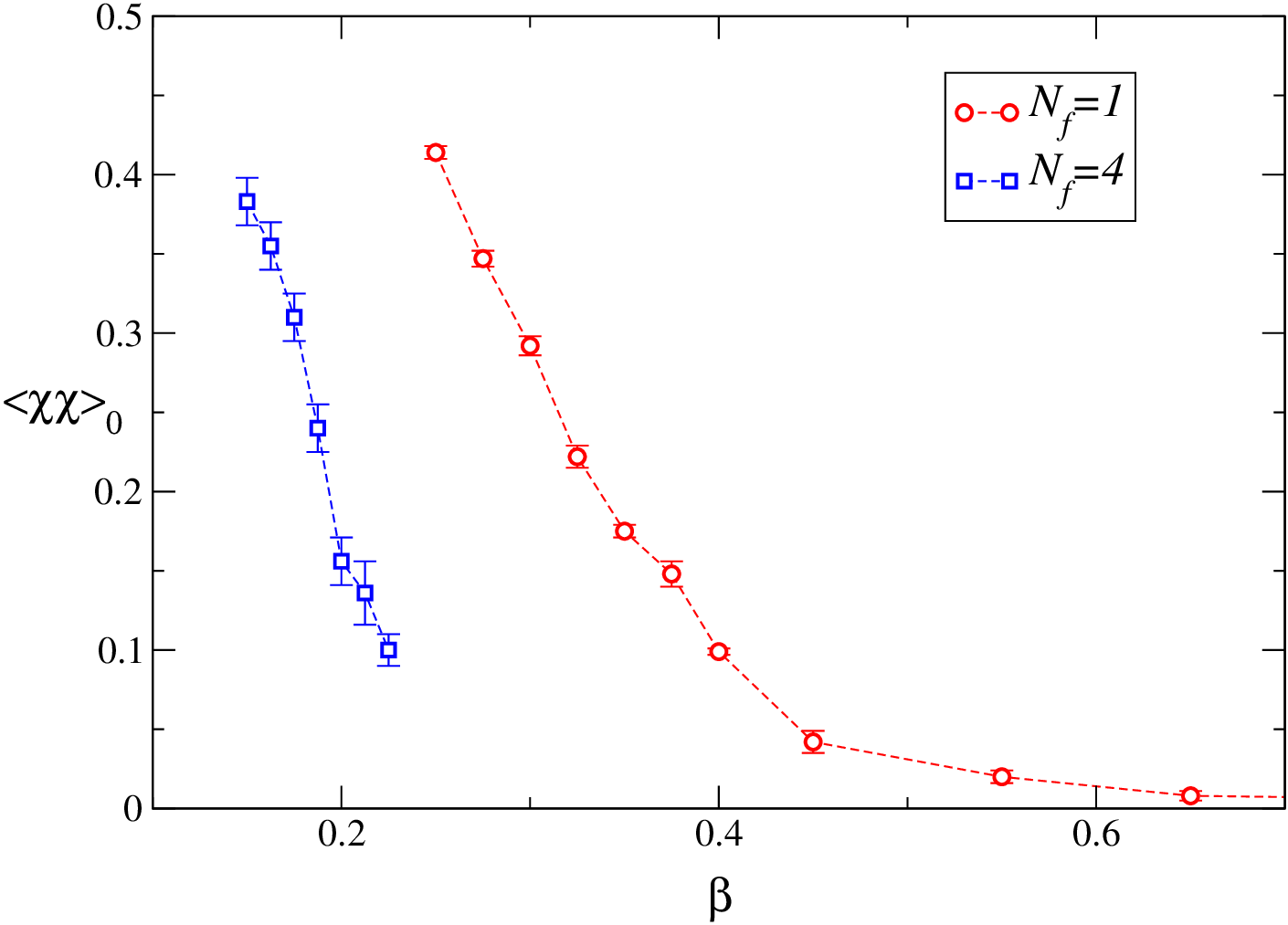}}

\smallskip
\caption[]{The value of $\langle\bar\chi\chi\rangle$ obtained in the chiral 
limit vs. $\beta$}
\label{fig:chiral}
\end{figure}

In an attempt to get an overview of the model's behaviour, for each $\beta$
and $N_f$ we first attempt an
extrapolation to the chiral limit of the form
\begin{equation}
\langle\bar\chi\chi\rangle=\langle\bar\chi\chi\rangle_0+Am_0+Bm_0^{3\over2}
\end{equation}
and plot the resulting $\langle\bar\chi\chi\rangle_0$ as a function of $\beta$
in Fig.~\ref{fig:chiral}. 
We have checked that at least at stronger couplings where the
signal is appreciable the results
are relatively insensitive to changes in the 
extrapolation function, eg. to incorporate non-analyticities due to 
Goldstone degrees of freedom \cite{Engels}.
We see that for both $N_f$ values chiral symmetry clearly appears to be broken
at strong coupling, but that $\langle\bar\chi\chi\rangle_0$
decreases as $\beta$ is increased, so that we can
identify either a crossover or a symmetry restoring phase transition at 
$\beta\simeq0.5$ ($N_f=1$) or $\beta\simeq0.25$ ($N_f=4$). Global fits to the 
data will be used in Sec.~\ref{sec:global} to explore the nature of these points
further. There is some evidence already from Fig.~\ref{fig:chiral} that 
the curvature of $\langle\bar\chi\chi\rangle_0(\beta)$ differs between the two
values of $N_f$, a feature already noted in \cite{DKK}. 

\begin{figure}[p]

                \centerline{ \epsfysize=3.6in
                             \epsfbox{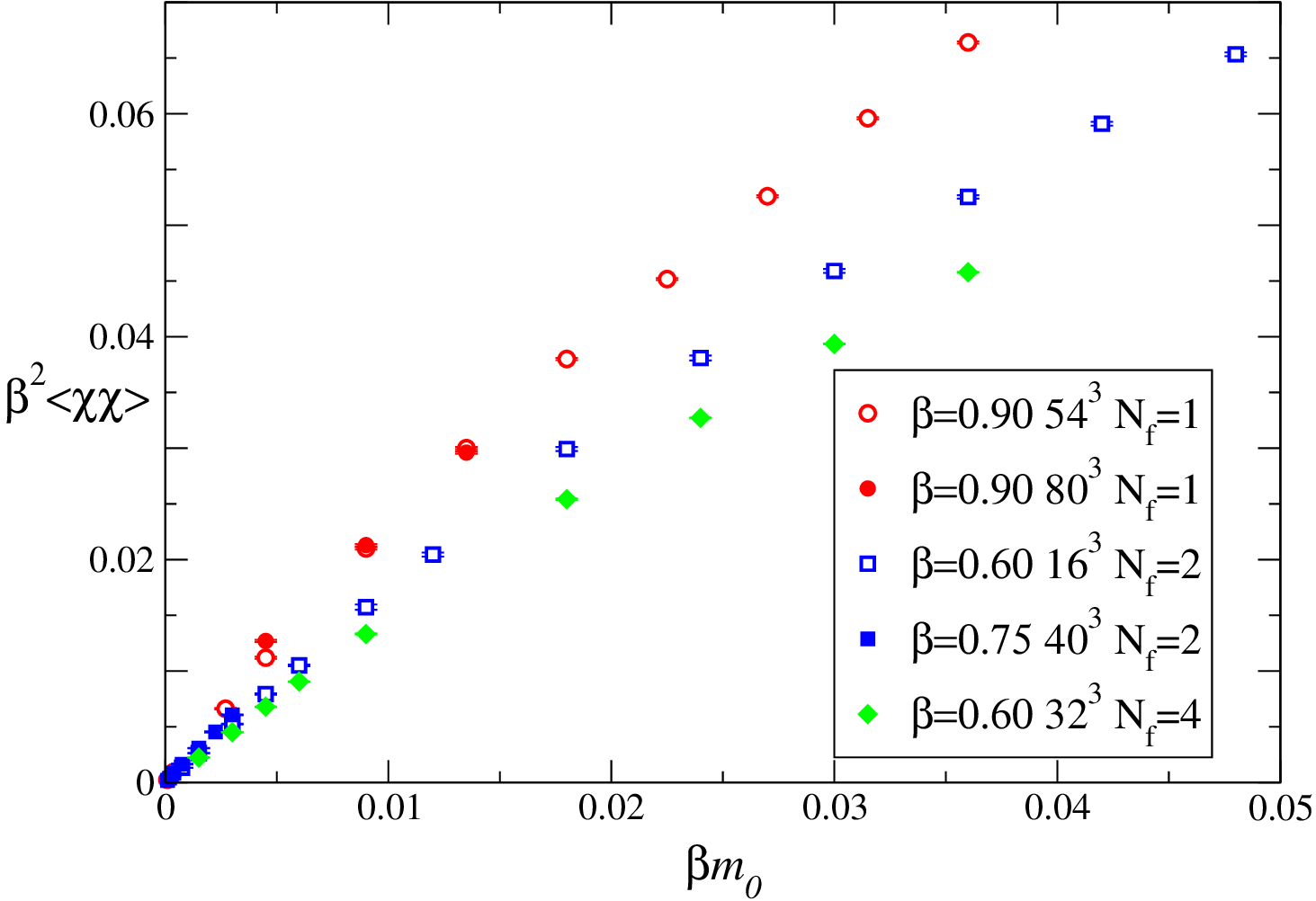}}

\smallskip
\caption[]{$\beta^2 \langle \bar\chi\chi \rangle$ vs. $\beta m_0$
 for $N_f=1,2$ and 4 for the datasets closest to continuum and chiral limits.}
\label{fig:continuum}
\end{figure}

In an attempt to see whether chiral symmetry is broken in the continuum limit
$\beta\to\infty$
we compare plots of the dimensionless variables
$\beta^2\langle\bar\chi\chi\rangle$ versus $\beta m_0$ for different $N_f$
in Fig.~\ref{fig:continuum}. The data for $N_f=2$ are from \cite{HKS}, where 
it was concluded that when extrapolated to the continuum limit
$\beta^2\langle\bar\chi\chi\rangle\lapprox10^{-4}$. By comparing $N_f=1$,
2 and 4 one sees that increasing $N_f$ at fixed $\beta m_0$
suppresses the condensate, as might be expected due to enhanced screening from
virtual $\chi\bar\chi$ pairs. There is clearly no prospect of finding a signal
for chiral symmetry breaking with $N_f=4$ with this dataset. The $N_f=1$ data,
however, are significantly larger near the chiral limit.
 
\begin{figure}[t]

                \centerline{ \epsfysize=3.6in
                             \epsfbox{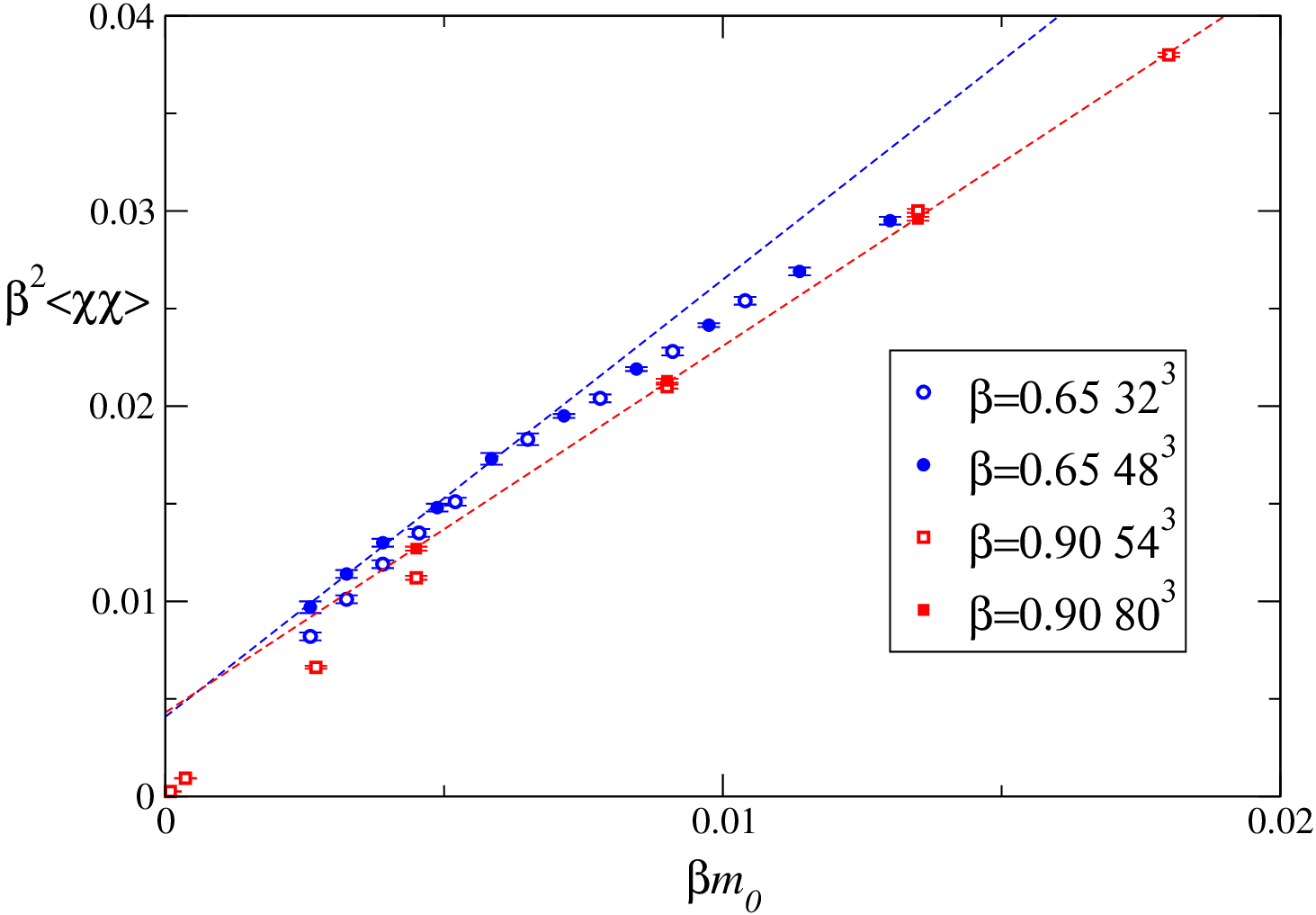}}

\smallskip
\caption[]{$\beta^2 \langle \bar\chi\chi \rangle$ vs. $\beta m_0$
for $N_f=1$ in the region of the chiral limit.}
\label{fig:continuum_zoom}
\end{figure}

In Fig.~\ref{fig:continuum_zoom} we show the plot for $N_f=1$ zoomed in the
neigbourhood of the origin for two different lattice spacings each with two
volumes. The smaller of the two $\beta=0.90$ lattices is intermediate 
in physical volume between the two $\beta=0.65$ lattices, whereas the
$\beta=0.90$ $80^3$
data is closest to both continuum and thermodynamic limits. There are clear
finite volume effects at both lattice spacings for $\beta m_0<0.005$. By
comparing $\beta=0.65$ with $\beta=0.90$ it is clear there are also still
lattice
discretisation artifacts of $O$(10\%). A linear fit to data on the larger
volume extrapolates to 
a chiral limit value $\beta^2\langle\bar\chi\chi\rangle=0.0041(4)$ for
$\beta=0.65$ (fit to the lowest 5 mass points), and 0.0043(2) for $\beta=0.90$,
with the fitted slopes 
differing by about 20\%\footnote{Note that the slope of the line
corresponds to the longitudinal susceptibility $\chi_l=\partial^2\ln Z/\partial
m_0^2$, which as a two-point function is expected to be more severely
finite-volume
affected.}. Due to finite volume effects 
these numbers are if anything underestimates.
Therefore whilst it is premature to quote a value for the
dimensionless condensate due to lack of control over systematic errors, we are
confident in claiming that it is 
$\gapprox10^{-3}$, ie. greater than zero, 
almost two orders of magnitude greater than the 
upper bound quoted for
$N_f=2$ in \cite{HKS}, and very roughly half the result 
for $N_f=0$ (ie. quenched
QED$_3$) found in \cite{HK}. Note, moreover, that our lowest mass datapoint 
at $\beta=0.90$ has $\beta^2\langle\bar\chi\chi\rangle$ only three times its 
value in the chiral limit, making the linear extrapolation of
Fig.~\ref{fig:continuum_zoom} 
considerably less adventurous than those of \cite{HK}.
Any extrapolation which obtains a significantly smaller
result in the chiral limit must
be based on an analytic form which curves sharply as the chiral limit is
approached. One possible form is $\langle\bar\chi\chi\rangle=a_0+
a_1 m_0^{1\over2}+\cdots$, predicted by the effective chiral theory describing the
influence of pion fluctuations (see eg. \cite{Engels}).
The applicability of such a model, however, presupposes chiral symmetry
breaking. Our conclusion that $N_{fc}>1$ is probably quite robust.

\subsection{Meson Spectroscopy}
\label{sec:pions}
A physically meaningful way of assessing how far the simulations lie from the
continuum limit comes by checking the degree of flavor symmetry restoration in
the meson spectrum. Recall that in the continuum theory 
with $N_f=2$ \cite{gusynin}, 
chiral symmetry breaking U(4) $\to$
U(2)$\otimes$U(2) results in eight Goldstone bosons, which somewhat sloppily 
we will continue to
refer to as ``pions''. If we denote the Euclidean time direction as $x_3$, then 
the pion states can be divided into 4 pseudoscalars 
$\bar\psi(\gamma_5\otimes\One)\psi$, $\bar\psi(\gamma_5\otimes\tau_i)\psi$
and 4 scalars 
$\bar\psi(\gamma_4\otimes\One)\psi$, $\bar\psi(\gamma_4\otimes\tau_i)\psi$,
where $\{\One_2,\vec\tau\}$ generate a U(2) symmetry acting on flavor indices. 
In the
lattice model the corresponding states in terms of staggered fermions read
\cite{burden}:
\begin{eqnarray}
\bar\psi(\gamma_5\otimes\One)\psi &\Leftrightarrow&
\bar\chi(x)\chi(x)\varepsilon(x); \label{eq:exact}\\
\bar\psi(\gamma_4\otimes\tau_\mu)\psi &\Leftrightarrow&
\bar\chi(x)\chi(x+\hat\mu)\varepsilon(x)\zeta_\mu(x); \label{eq:1link}\\
\bar\psi(\gamma_5\otimes\tau_\mu)\psi &\Leftrightarrow&
\vert\epsilon_{\mu\nu\lambda}\vert
\bar\chi(x)\chi(x+\hat\nu+\hat\lambda)\varepsilon(x)\zeta_\nu(x)\zeta_\lambda(x);
 \label{eq:2link}\\
\bar\psi(\gamma_4\otimes\One)\psi &\Leftrightarrow&
\vert\epsilon_{\mu\nu\lambda}\vert
\bar\chi(x)\chi(x+\hat\mu+\hat\nu+\hat\lambda),
\end{eqnarray}
where we have defined additional phase factors
$\varepsilon(x)=(-1)^{x_1+x_2+x_3}$,
$\zeta_\mu(x)=\varepsilon(x)\eta_\mu(x)(-1)^{x_\mu}$, and an average over forward
and backward link shifts is understood. The parity operation for
staggered fermions with the definition (\ref{eq:KS}) for the Kawamoto-Smit
phases is
\begin{equation}
x=(x_1,x_2,x_3)\mapsto x^\prime=(1-x_1,x_2,x_3)\;\;\;;\;\;\;
\chi(x)\mapsto\varepsilon(x^\prime)\chi(x^\prime)\;,\;
\bar\chi(x)\mapsto\varepsilon(x^\prime)\bar\chi(x^\prime)
\end{equation}
The assignment of scalar and pseudoscalar labels to the states is
readily checked. 

Spectroscopy on the lattice through decay of temporal correlation functions
is most transparent when restricted to operators 
defined on a single timeslice. This means that in addition to the state
(\ref{eq:exact}) which interpolates an exact Goldstone state for the residual
lattice chiral symmetry breaking U(1)$\otimes$U(1)$\to$ U(1), it is possible to
examine two scalar states interpolated by one-link operators (\ref{eq:1link})
and one  pseudoscalar interpolated by a two-link operator (\ref{eq:2link}).
At non-zero lattice spacing these latter states are not associated with
Goldstones of an exact symmetry, and the degeneracy between the states expected
in the continuum limit is split by terms which are generically $O(a)$. The
extent to which the measured masses are degenerate, therefore, gives some
measure of the approach to the continuum limit. Note that flavor symmetry
restoration should still manifest itself in this fashion even if chiral symmetry 
is unbroken, since in this case the various states are related by
U(2)$\otimes$U(2)
rotations.

\begin{figure}[p]

                \centerline{ \epsfysize=4.2in
                             \epsfbox{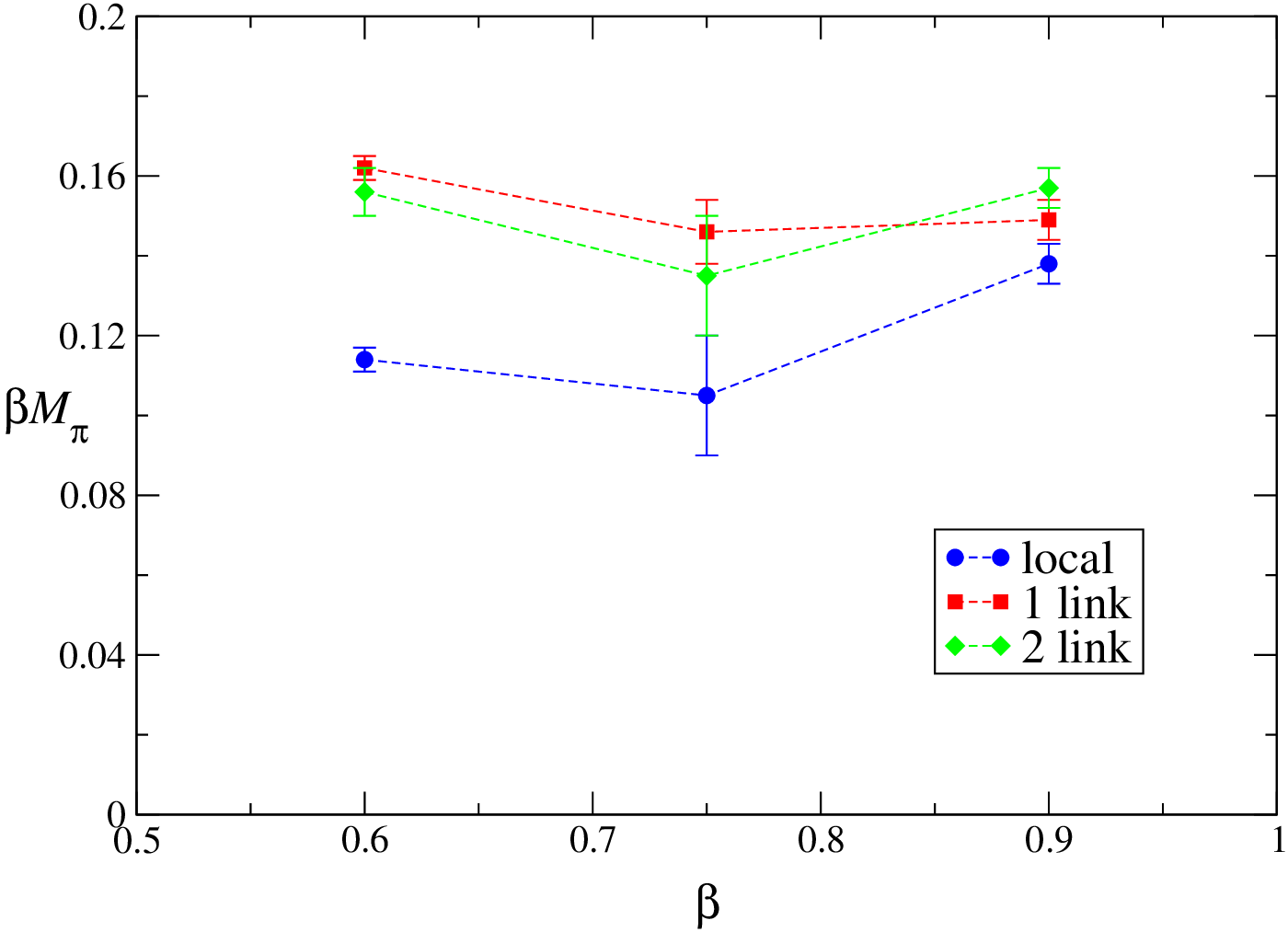}}

\smallskip
\caption[]{Dimensionless masses $\beta M_\pi$ vs. $\beta$ for the
various pion operators discussed in Sec.~\ref{sec:pions}.}
\label{fig:pions}
\end{figure}

We have tested the approach to the continuum limit for $N_f=1$
by measuring the masses of
the local, 1-link and 2-link pions at $\beta=0.6$, 0.75 and 0.9 with fixed
values of $\beta m_0=0.0009$ and $L_s/\beta=40$. In this way any systematic
errors due to eg. finite volume should cancel. The results are plotted in
Fig.~\ref{fig:pions}. Whilst the data is still too noisy to 
determine the level ordering unequivocally, 
we can see that the splitting of 40\% at
$\beta=0.6$ is reduced to 15\% by $\beta=0.9$, which thus gives a sample
estimate for the residual systematic errors due to lattice artifacts at this 
closest approach to the continnuum limit.

\subsection{Global Fits to Equation of State}
\label{sec:global}

\begin{figure}[p]

                \centerline{ \epsfysize=4.2in
                             \epsfbox{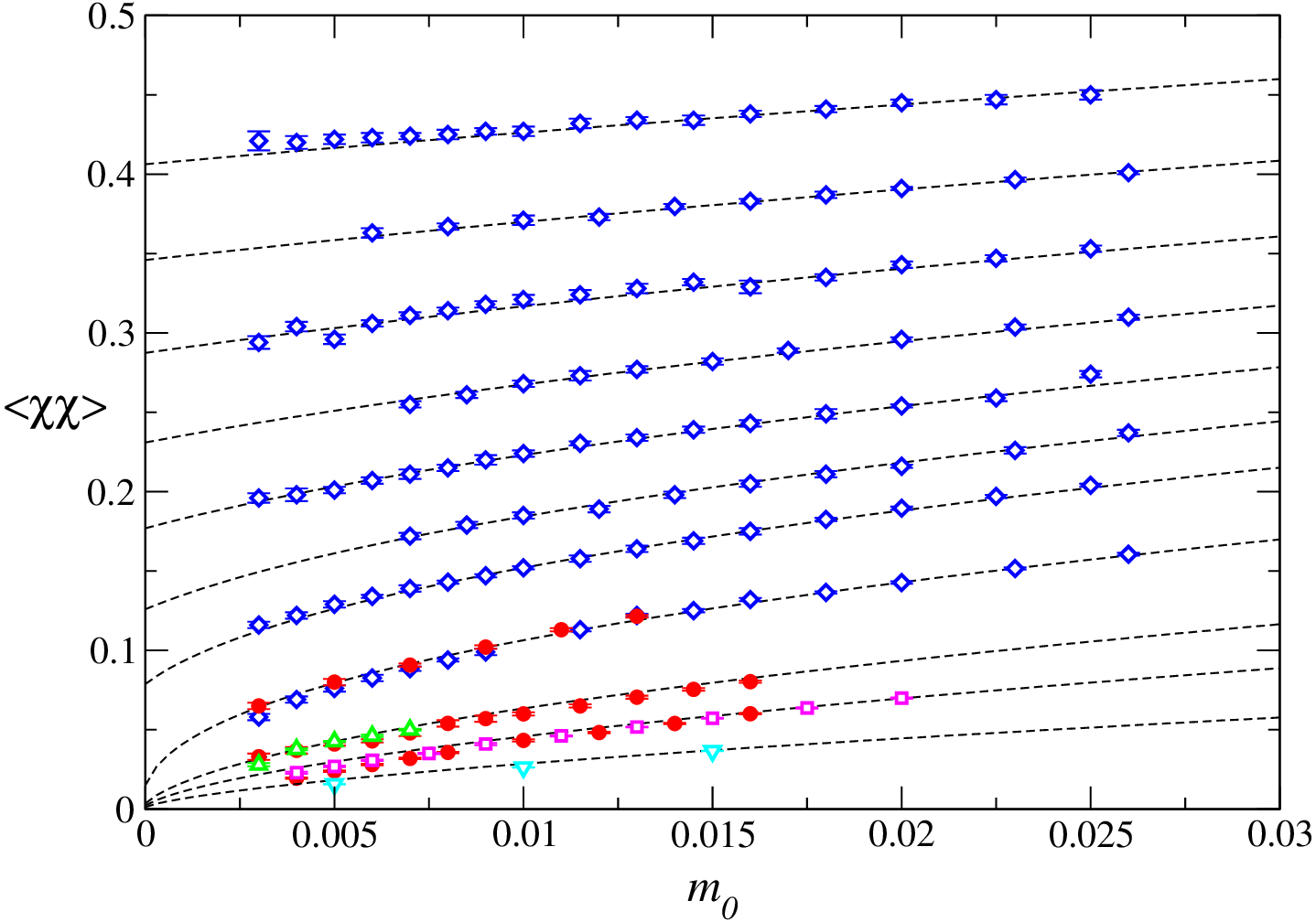}}

\smallskip
\caption[]{Chiral condensate $ \langle\bar\chi\chi\rangle$
vs. mass $m_0$ for $N_f=1$ from $24^3$ (diamonds), $32^3$ (filled circles), $44^3$
(triangles), $48^3$ (squares) and $80^3$ ($\nabla$)
lattices. The lines denote fits to data at 
constant $\beta$ to be discussed in Sec.~\ref{sec:global}, for $\beta$ values
(from the top) 0.25, 0.275, 0.3, 0.325, 0.35, 0.375, 0.4, 0.45, 0.55, 0.65
and 0.90.}
\label{fig:gplot1}
\end{figure}

\begin{figure}[p]

                \centerline{ \epsfysize=4.2in
                             \epsfbox{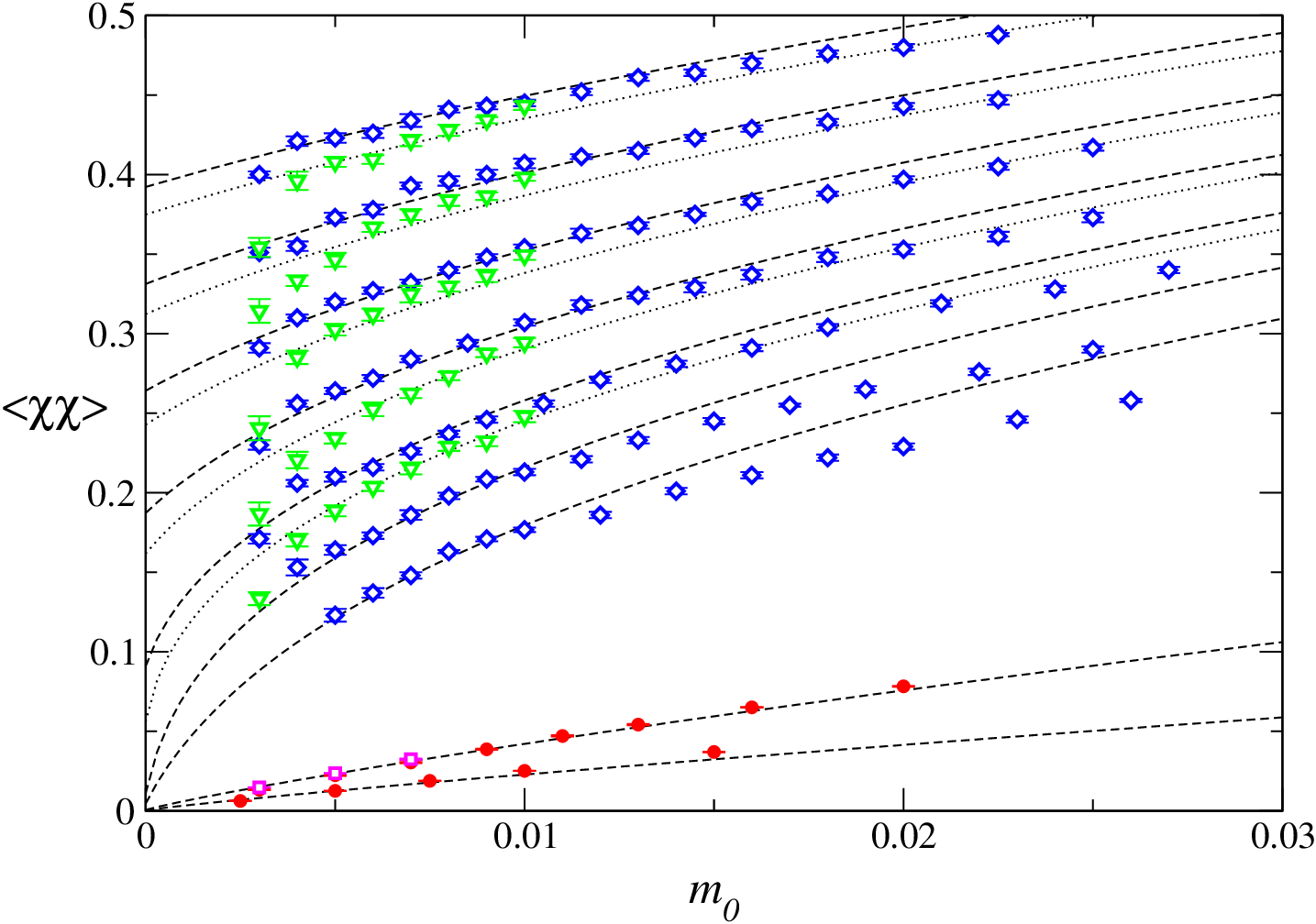}}

\smallskip
\caption[]{$\langle\bar\chi\chi\rangle$ vs. $m_0$ for $N_f=4$
from $16^3$ (triangles), $24^3$ (diamonds), $32^3$ (filled circles) and 
$48^3$ (squares) lattices. The lines denote finite volume scaling fits to data
at constant $\beta$ (see Sec.~\ref{sec:global}), for $\beta$ values (from the
top) 0.15, 0.1625, 0.175, 0.1875, 0.2, 0.2125, 0.225, 0.4 and 0.6, with dotted
lines denoting fits to $16^3$ and dashed lines fits to all other volumes.}
\label{fig:gplot4}
\end{figure}

In this section we change tack, retreating from direct attempts to probe the
continuum limit, and instead exploiting the data at stronger couplings to make
qualitative statements about the behaviour as a function of $N_f$. In the
strong coupling limit $\beta\to0$, it is known rigorously that chiral
symmetry is spontaneously broken \cite{SS}, and indeed our data confirm this
(see Figs.~\ref{fig:gplot1},\ref{fig:gplot4}). Therefore as $\beta$ increases,
for $N_f>N_{fc}$ there must be a chiral symmetry restoring phase transition for
some finite $\beta_c$. For $N_f<N_{fc}$, since we believe the order parameter is
exponentially small in the continuum limit, the relic of this transition may 
persist as a crossover between strong and weak coupling behaviour, much as in
lattice QCD. For $N_f$ exactly equal to $N_{fc}$, the order parameter should 
vanish
in the chiral continuum limit, but with a non-analytic response to a small
bare mass of the form $\langle\bar\chi\chi\rangle\propto m_0^\delta$, a
behaviour characteristic of a so-called ``conformal fixed point''
\cite{conformal, gusynin}.

We therefore proceed with the assumption (which must be unphysical for
$N_f<N_{fc}$) that there is a continuous chiral symmetry-restoring phase
transition described by the following equation of state:
\begin{equation}
m_0=A(\beta-\beta_c)\langle\bar\chi\chi\rangle^p
+B\langle\bar\chi\chi\rangle^\delta,
\label{eq:eos}
\end{equation}
and attempt to fit the entire dataset, or at least a significant fraction, for
each $N_f$. Eqn.~\ref{eq:eos} has 5 free parameters, the ``critical'' coupling
$\beta_c$, exponents $\delta$ and $p$, and the amplitudes $A$ and $B$. 
Setting $\beta=\beta_c$ shows that $\delta$ has its conventional interpretation
as a critical exponent describing response of an order parameter to an external
field, and setting $m_0=0$ shows that the conventional exponent describing
scaling of the order parameter in the broken symmetry phase is given by 
$\beta_{mag}=(\delta-p)^{-1}$. The fits are performed using the MINUIT
package.

For $N_f=1$ there are a total of 151 points in the dataset (excluding
$\beta=0.9$). The best
fit found was to the 109 points with $0.25\leq\beta\leq0.45$ 
using data from all masses
and from the
largest available volume at any given $\beta$. This gives
\begin{equation}
\begin{array}{lllll}
A=4.43(17) & \delta=2.33(4) & \beta_c=0.451(4) &&\chi^2/dof=1.02 \\
B=1.87(7)  & p=1.50(2) &&&\\
\end{array}
\label{eq:fit1}
\end{equation}
Acceptable fits persist, with little change in the parameters, if data with
$\beta=0.400,0.450$ is excluded. The $\chi^2/dof$ increases to 1.5 if
$\beta=0.55$ is included, and 2.5 if $\beta=0.65$ is included, once again with
relatively little impact on the fit parameters. If instead 
strong coupling data with 
$\beta\leq0.40$ are excluded, then $\chi^2/dof$ rises to $\gapprox4$, and the
fitted values of $\delta$ and $p$ drift downwards.

\begin{figure}[t]

                \centerline{ \epsfysize=3.6in
                             \epsfbox{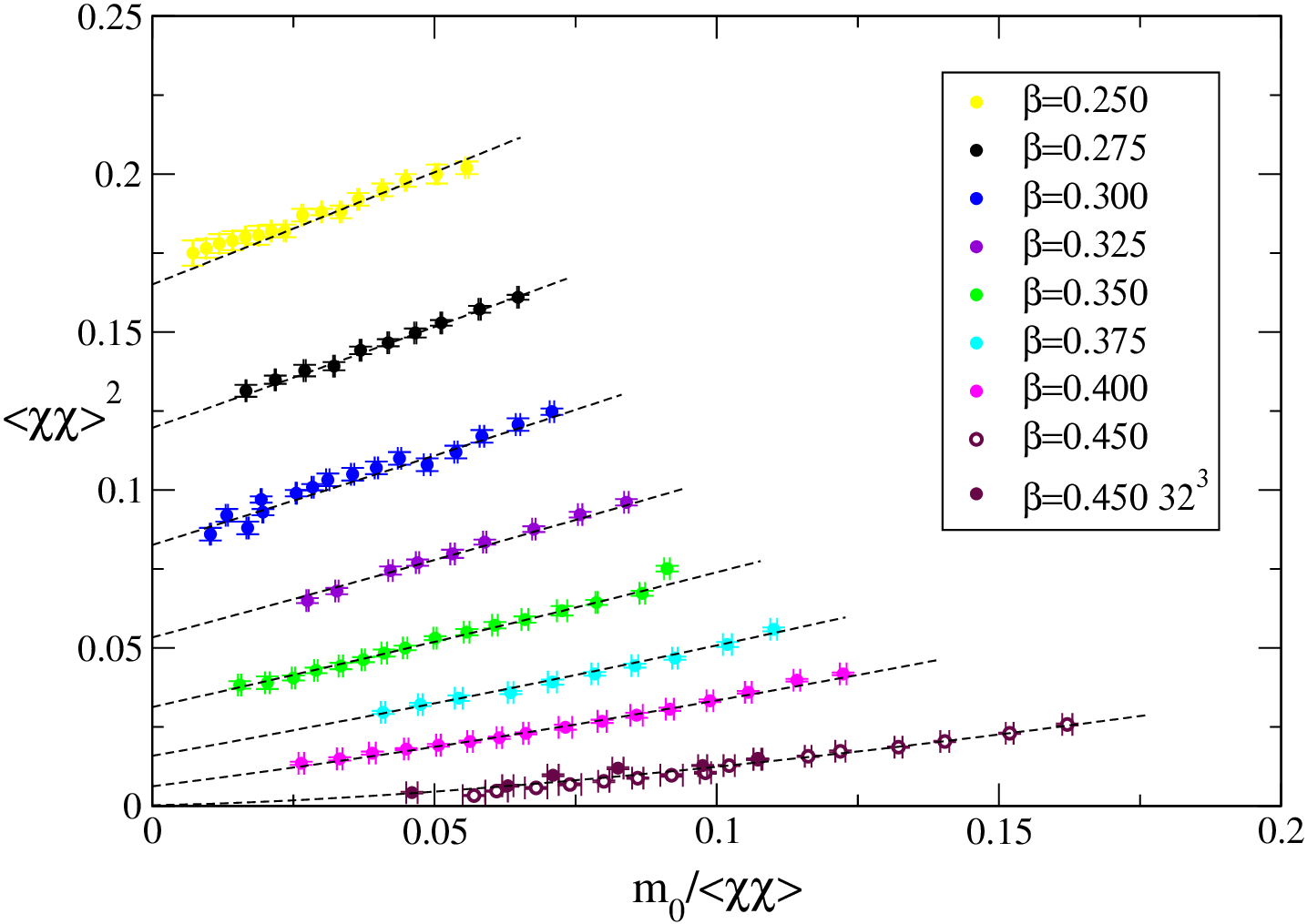}}

\smallskip
\caption[]{Fisher plot $\langle\bar\chi\chi\rangle^2$ vs.
$m_0/\langle\bar\chi\chi\rangle$ for strong coupling data with $N_f=1$.
Unless stated all data are from $24^3$ lattices.
The dashed lines are the result of the fit (\ref{eq:fit1}).}
\label{fig:fishfit1_sc}
\end{figure}

\begin{figure}[p]

                \centerline{ \epsfysize=3.6in
                             \epsfbox{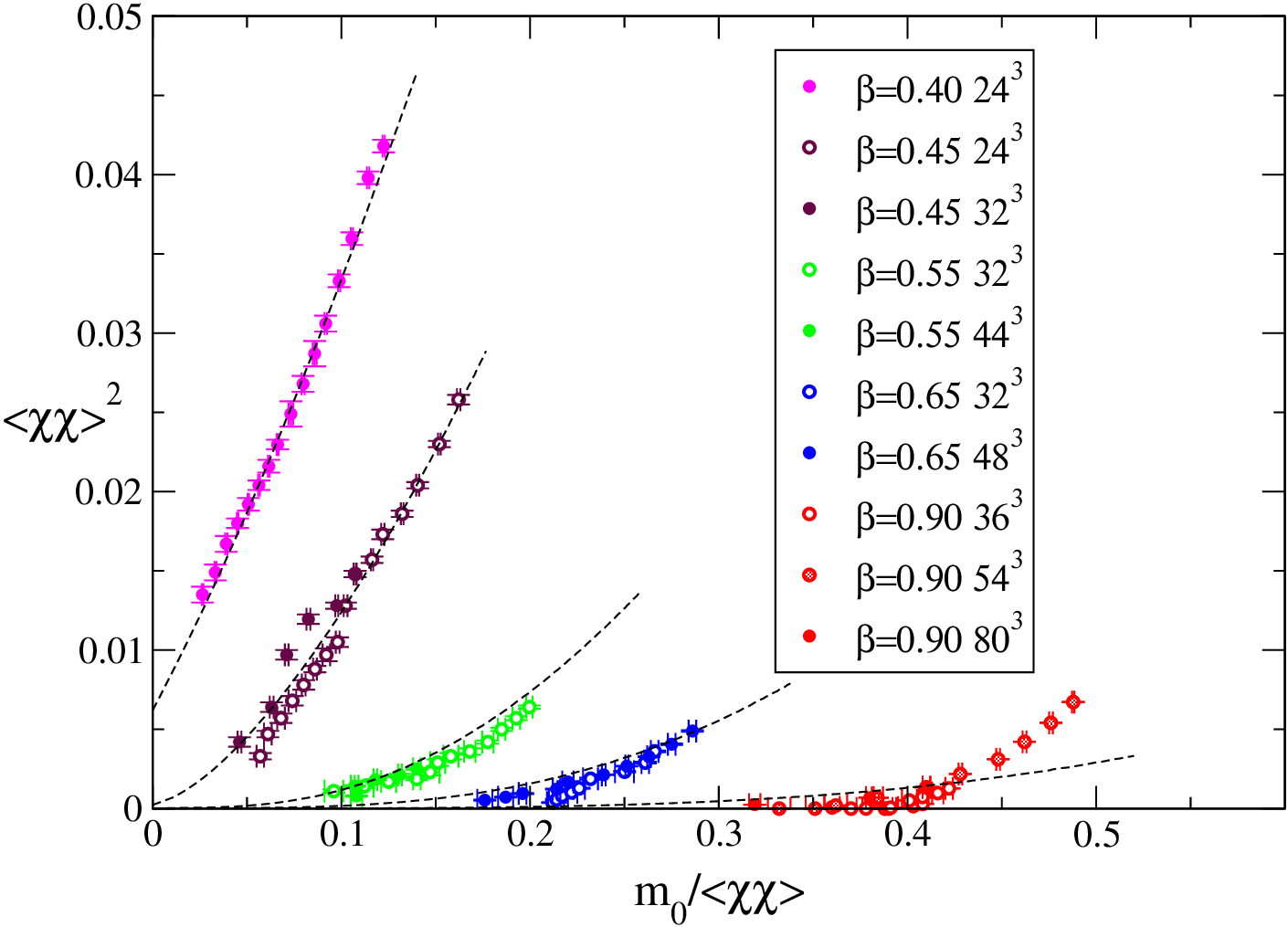}}

\smallskip
\caption[]{Fisher plot for weak coupling data with $N_f=1$.}
\label{fig:fishfit1_wc}
\end{figure}

The raw data are plotted with
the fit (\ref{eq:fit1}) in Fig.~\ref{fig:gplot1}, and the fit is also shown in
``Fisher coordinates'' for strong (Fig.~\ref{fig:fishfit1_sc}) and weak
(Fig.~\ref{fig:fishfit1_wc}) coupling data. The Fisher plot of
$\langle\bar\chi\chi\rangle^2$ 
versus $m_0/\langle\bar\chi\chi\rangle$ is devised so
that data described by a Landau-Ginzburg equation of state (ie. $\delta=3$ and
$p=1$ in (\ref{eq:eos})) yield trajectories of constant $\beta$ as parallel
straight lines, intersecting the $y$-axis for $\beta<\beta_c$, the $x$-axis for 
$\beta>\beta_c$, and the origin for $\beta=\beta_c$.

Overall, the quality of the fit is remarkably good. 
The only sign of inconsistency is the disagreement
in the shape of the curve in the weak coupling phase with
$\beta\geq0.55$, although even in this regime the fit apparently passes close to
``centre of mass'' of the data. As seen in Fig.~\ref{fig:fishfit1_wc}, the data
here lie below the fit but curve more strongly, which they must do if 
they are ultimately to
intercept the $y$-axis in the chiral limit.

\begin{figure}[p]

                \centerline{ \epsfysize=3.6in
                             \epsfbox{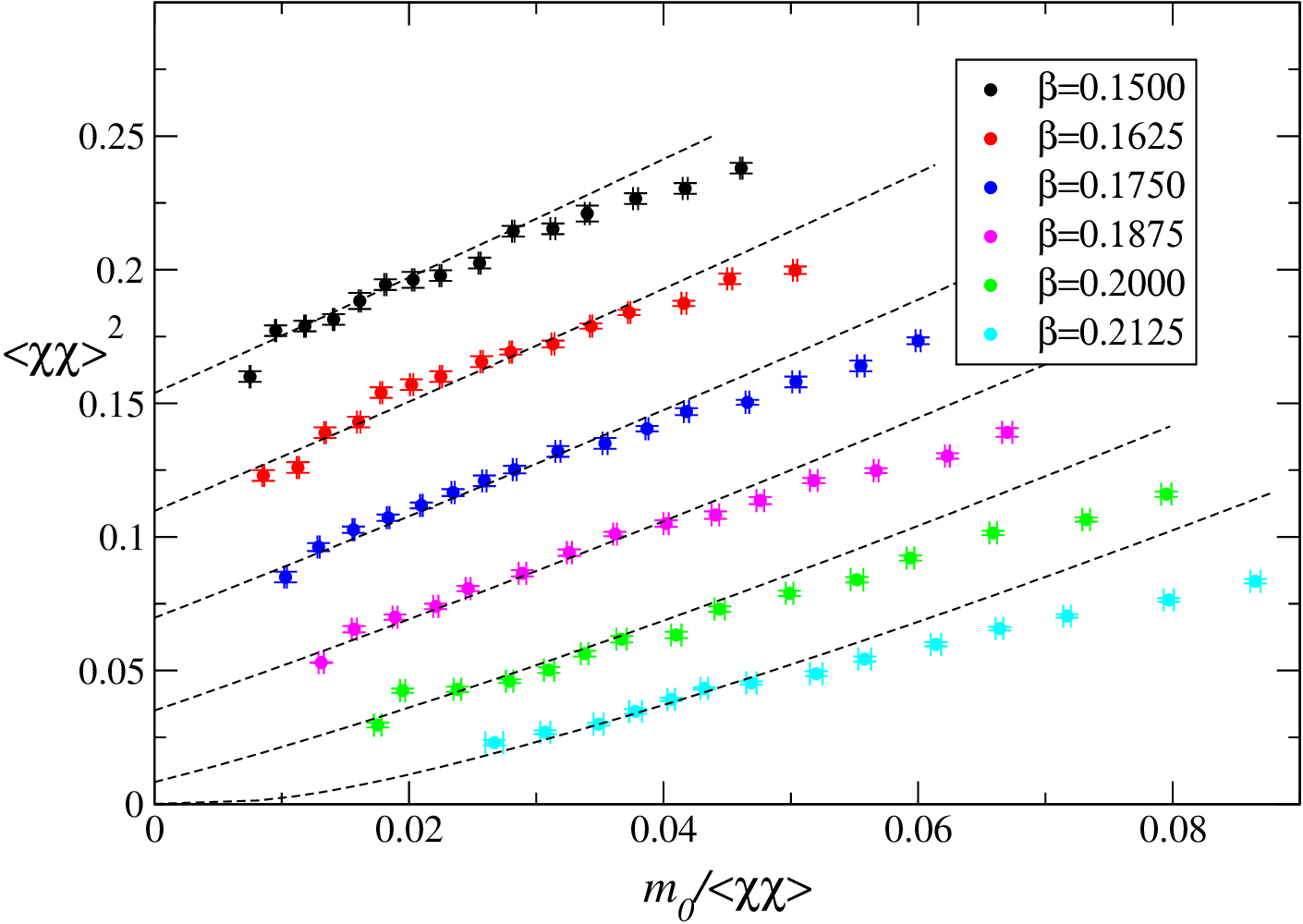}}

\smallskip
\caption[]{Fisher plot of the strong coupling data from a $24^3$ lattice for
$N_f=4$; the dashed lines are the result of the fit (\ref{eq:fit4fvs}).}
\label{fig:fishfit4_sc}
\end{figure}

\begin{figure}[p]

                \centerline{ \epsfysize=3.6in
                             \epsfbox{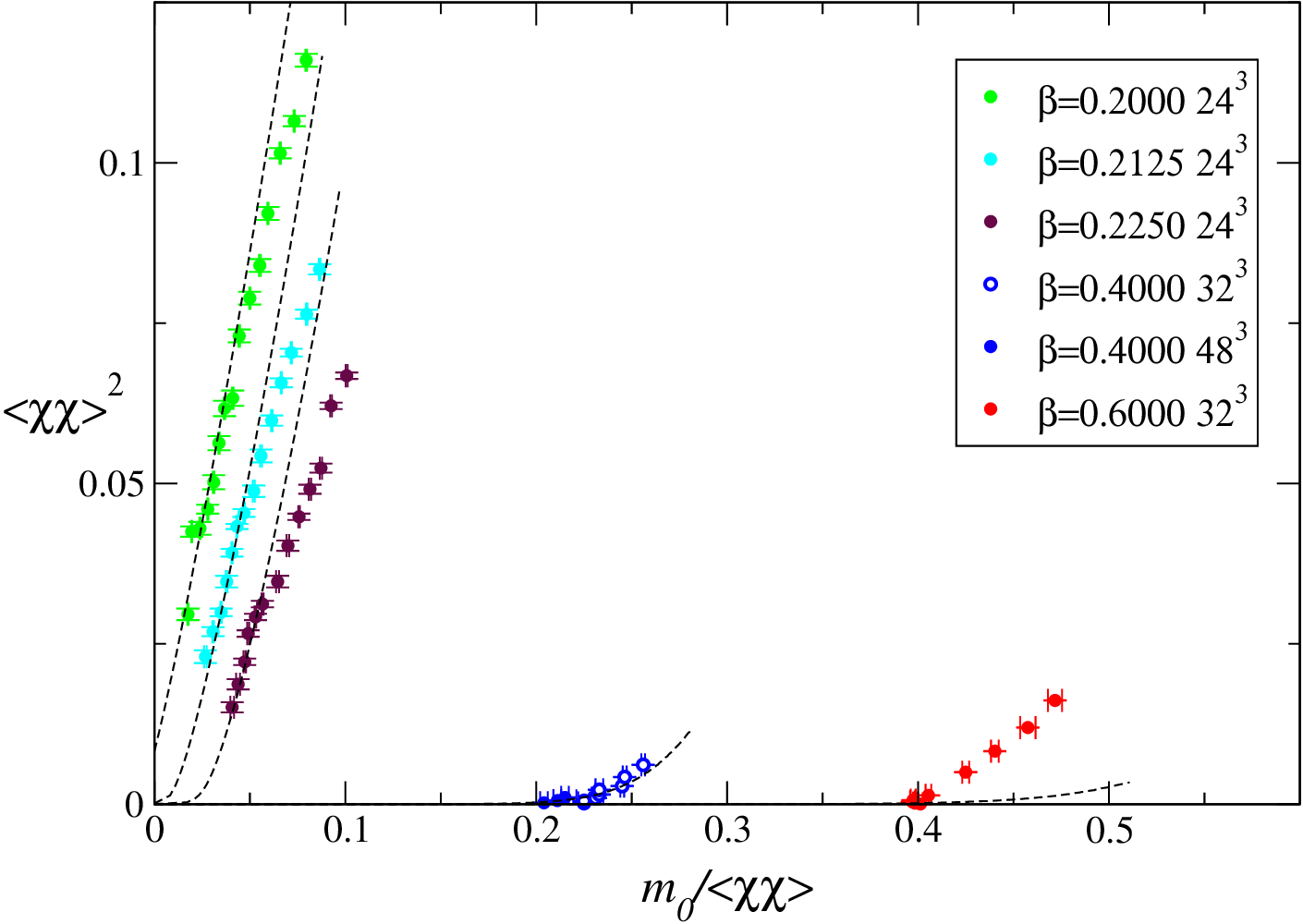}}

\smallskip
\caption[]{Fisher plot of the weak coupling data for $N_f=4$.}
\label{fig:fishfit4_wc}
\end{figure}

For $N_f=4$, we were unable to find any fit of comparable quality. Global fits
to all 143 points in the dataset typically yielded a $\chi^2/dof\simeq O(5)$. 
Since in Fig.~\ref{fig:gplot4}
the low mass points are clearly finite volume effected, one strategy is to
exclude them and only fit to the 42 datapoints with $\beta\leq0.20$,
$m_0\geq0.01$, yielding
\begin{equation}
\begin{array}{lllll}
A=2.27(29) & \delta=2.59(18) & \beta_c=0.212(4) &&\chi^2/dof=0.95 \\
B=0.60(7)  & p=0.98(13) &&&\\
\end{array}
\label{eq:fit4}
\end{equation}
A slightly more satisfactory fit including a wider window in $\beta$ was found
when some attempt at incorporating a finite-volume scaling  analysis was made, 
by modifying the equation of state to \cite{Thirring1}
\begin{equation}
m_0=A((\beta-\beta_c)+CL_s^{-{1\over\nu}})\langle\bar\chi\chi\rangle^p+
B\langle\bar\chi\chi\rangle^\delta,
\label{eq:eosfvs}
\end{equation}
where for the exponent $\nu$ we use the hyperscaling prediction
\begin{equation}
\nu={{(\delta+1)}\over{3(\delta-1)}}.
\end{equation}
The best fit to (\ref{eq:eosfvs}) was found for the the 121 datapoints with 
$0.15\leq\beta\leq0.4$, $m_0\leq0.015$:
\begin{equation}
\begin{array}{lllll}
A=1.988(24) & \delta=2.653(30) & \beta_c=0.2129(8) &&\chi^2/dof=8.1 \\
B=0.451(9)  & p=1.152(3) & C=0.176(7)&&\\
\end{array}
\label{eq:fit4fvs}
\end{equation}
Although the $\chi^2/dof$ is much higher, this fit does a better job in the 
``symmetric'' phase $\beta\geq0.4$, and is shown by dashed lines for in
Fig.~\ref{fig:gplot4}, and in Fisher coordinates in Fig.~\ref{fig:fishfit4_sc}
(strong coupling) and Fig.~\ref{fig:fishfit4_wc} (weak coupling). 
In any case,
the fitted values of the critical parameters do not vary much between
(\ref{eq:fit4}) and (\ref{eq:fit4fvs}).
 
As for 
$N_f=1$, the Fisher plot 
highlights systematic discrepancies in the symmetric
phase. Overall, though, it is difficult to conclude on the basis of this
analysis whether chiral symmetry is restored for $N_f=4$
as $\beta\to\infty$ or not. There
is, however, a significant change in the fitted exponent 
$\beta_{mag}$ as $N_f$ changes from 1, where from (\ref{eq:fit1}) it has the
value 1.20(3), to $N_f=4$ where it takes the value 0.62(9) (\ref{eq:fit4})
or 0.67(1) (\ref{eq:fit4fvs}). This trend is consistent with the
distinct curvatures of the $\langle\bar\chi\chi\rangle_0(\beta)$ data in
Fig.~\ref{fig:chiral}, and is a more dramatic effect than the increase
of $\delta(N_f)$, which even for $N_f=4$ still lies below the Landau-Ginzburg
value. 

\section{Discussion}

Our main result is that there is convincing evidence for
spontaneously broken chiral symmetry in the continuum limit
in the theory with $N_f=1$; any
extrapolation of the data of Fig.~\ref{fig:continuum_zoom} resulting in a
vanishing condensate in this limit must rely on a hitherto unobserved curvature
close to the chiral limit. The one theoretical framework in which we might
understand the origin of such a curvature, namely chiral effective theory, has
the assumption of chiral symmetry breaking built in. Therefore we can say
with some confidence that $N_{fc}>1$, and with slightly less confidence that 
the dimensionless condensate $\beta^2\langle\bar\psi\psi\rangle\sim O(10^{-3})$.

For $N_f=4$ the situation is much less clear cut.  We could not
possibly hope to confirm $N_{fc}>4$ by direct simulation using current
resources, since even for $N_f=2$ the dimensionless condensate, if there,
is too small to measure \cite{HKS}. We have instead contrasted $N_f=4$
with $N_f=1$ by treating the passage from strong to weak coupling as if there
were a true phase transition, and attempted to characterise the equation of
state by critical exponents. It is somewhat ironic that this proved quite
successful for $N_f=1$ where we believe there is no transition but merely a
crossover; indeed the quality of the fit is 
if anything superior to 
similar fits obtained in systems where a true phase transition is believed to
occur (eg. \cite{Thirring1,NJL4,Thirring2}). For $N_f=4$ the fits are less
compelling, although this could merely reflect a less
comprehensive dataset. The only
significant difference between the two theories is in the exponent
$\beta_{mag}$, which halves as $N_f$ increases from 1 to 4. There is also a 
small increase in $\delta$ from 2.3 to 2.7, which should  
be contrasted with the chiral symmetry restoring
transition in the 2+1$d$ Thirring model (which has the same global symmetries
as QED$_3$ in both lattice and continuum theories), 
which for $N_f=2$ has $\delta\simeq2.7$, increasing 
to $3.1$ for $N_f=3$, $3.4$ for $N_f=4$, and finally 
to $\gapprox4$ for $N_f=5$ \cite{Thirring1, Thirring2,Thirring3} 
(data for the Thirring
model with $N_f=1$ do not exist). Of course, the hypothesis that the two models
have a common fixed point behaviour implying the exponents should coincide only
holds for $N_f=N_{fc}$. Naively one might deduce from the trends in these
numbers that this would occur for $N_f<2$; however, one should bear in mind
that the equivalence should only hold for Thirring transitions at $G^2=\infty$
and QED transitions at $\beta=\infty$. 
In the Thirring case there is some ambiguity
about identifying the strong coupling limit using a lattice regularisation
\cite{Thirring1}, and 
it is unlikely that such information about QED could ever emerge from a
numerical simulation, though of course, it may be an interesting question to
address for 
analytic approaches. It is also interesting to speculate about the nature of the
universality class of the transition at finite $\beta_c$ for $N_f>N_{fc}$, in 
particular the question of whether it is 
governed by the global symmetries of the continuum or the lattice model.

\section*{Acknowledgements}
SJH was supported by a PPARC Senior Research Fellowship, and 
thanks the Institute for Nuclear Theory at the University of Washington for its
hospitality and the Department of Energy for partial support during the
completion of this work.
JBK is supported in part by NSF grant PHY-0102409. 
The computer simulations were done on the Cray SV1's at NERSC, the IBM-SP
at NPACI, 
the SGI Origin 2000 at the University of Wales Swansea, and on PCs at 
DESY Hamburg and Humboldt Universit\"at Berlin.  
We have enjoyed discussions with Shailesh Chandrasekharan, Valery Gusynin,
Igor Herbut, Nick Mavromatos, Manuel Reenders,
Rohana Wijewardhana, Pieter Maris and Zlatko Te\v sanovi\'c.


\newpage

\newpage

\newpage

\newpage

\newpage

\newpage

\newpage

\begin{thebibliography}{10}

%
\bibitem{pisarski}
R.~D.~Pisarski,
Phys.\ Rev.\ D {\bf 29} (1984) 2423.
%
\bibitem{SD}
T.~Appelquist, D.~Nash and L.~C.~R.~Wijewardhana,
Phys.\ Rev.\ Lett.\  {\bf 60} (1988) 2575;\\
M.R. Pennington and S.R. Webb, Brookhaven preprint
BNL-40886;\\
M.~R.~Pennington and D.~Walsh,
Phys.\ Lett.\ B {\bf 253} (1991) 246;\\
P.~Maris,
Phys.\ Rev.\ D {\bf 54} (1996) 4049
[arXiv:hep-ph/9606214].
%
\bibitem{DKK}
E. Dagotto, A. Koci\'c and J.B. Kogut,
Phys. Rev. Lett. {\bf62} (1989) 1083; 
Nucl. Phys. {\bf B334} (1990) 279;
%
\bibitem{lattice}
V.~Azcoiti and X.~Q.~Luo,
Mod.\ Phys.\ Lett.\ A {\bf 8} (1993) 3635
[arXiv:hep-lat/9212011];\\
J.~Alexandre, K.~Farakos, S.~J.~Hands, G.~Koutsoumbas and S.~E.~Morrison,
Phys.\ Rev.\ D {\bf 64} (2001) 034502
[arXiv:hep-lat/0101011].
%
\bibitem{HK}
S.~J.~Hands and J.~B.~Kogut,
Nucl.\ Phys.\ B {\bf 335} (1990) 455.
%
\bibitem{HKS}
S.~J.~Hands, J.~B.~Kogut and C.~G.~Strouthos,
Nucl.\ Phys.\ B {\bf 645} (2002) 321
[arXiv:hep-lat/0208030].
%
\bibitem{AW}
T.~Appelquist and L.~C.~R.~Wijewardhana,
arXiv:hep-ph/0403250.
%
\bibitem{Pieter}
P. Maris, private communication.
%
\bibitem{Tesanovic}
M.~Franz, Z.~Te\v sanovi\'c and O.~Vafek, Phys.\ Rev.\ B{\bf66} (2002) 054535 
[arXiv:cond-mat/0203333].
%
\bibitem{Herbut}
I.~F.~Herbut, Phys.\ Rev.\ B{\bf66} (2002) 094504 [arXiv:cond-mat/0202491].
%
\bibitem{FT}
M.~Franz and Z.~Te\v sanovi\'c, Phys.\ Rev.\ Lett.\ {\bf84} (2000) 554.
%
\bibitem{Alex}
A.~Kovner, B.~Rosenstein and D.~Eliezer,
Mod.\ Phys.\ Lett.\ A {\bf 5} (1990) 2733.
%
\bibitem{appelquist}
T.~Appelquist, A.~G.~Cohen and M.~Schmaltz,
Phys.\ Rev.\ D {\bf 60} (1999) 045003
[arXiv:hep-th/9901109].
%
\bibitem{Mavro}
N.~E.~Mavromatos and J.~Papavassiliou,
arXiv:cond-mat/0311421.
%
\bibitem{Espriu}
D.~Espriu, A.~Palanques-Mestre, P.~Pascual and R.~Tarrach,
Z.\ Phys.\ C {\bf 13} (1982) 153;\\
S.~J.~Hands,
Phys.\ Rev.\ D {\bf 51} (1995) 5816
[arXiv:hep-th/9411016].
%
\bibitem{Itoh}
T.~Itoh, Y.~Kim, M.~Sugiura and K.~Yamawaki,
Prog.\ Theor.\ Phys.\  {\bf 93} (1995) 417
[arXiv:hep-th/9411201].
%
\bibitem{Thirring1}
L.~Del Debbio, S.~J.~Hands and J.~C.~Mehegan  [UKQCD Collaboration],
Nucl.\ Phys.\ B {\bf 502} (1997) 269
[arXiv:hep-lat/9701016].
%
\bibitem{Thirring2}
L.~Del Debbio and S.~J.~Hands,
Nucl.\ Phys.\ B {\bf 552} (1999) 339
[arXiv:hep-lat/9902014].
%
\bibitem{Thirring3}
S.~J.~Hands and B.~Lucini,
Phys.\ Lett.\ B {\bf 461} (1999) 263
[arXiv:hep-lat/9906008].
%
\bibitem{Gottlieb}
S.~Gottlieb, W.~Liu, D.~Toussaint, R.~L.~Renken and R.~L.~Sugar,
Phys.\ Rev.\ D {\bf 35} (1987) 2531.
%
\bibitem{Hasenfratz}
P.~Hasenfratz,
Nucl.\ Phys.\ Proc.\ Suppl.\  {\bf 106} (2002) 159
[arXiv:hep-lat/0111023].
%
\bibitem{GusyninReenders}
V.~P.~Gusynin and M.~Reenders,
Phys.\ Rev.\ D {\bf 68} (2003) 025017
[arXiv:hep-ph/0304302].
%
\bibitem{Engels}
J.~Engels and T.~Mendes,
Nucl.\ Phys.\ B {\bf 572} (2000) 289
[arXiv:hep-lat/9911028].
%
\bibitem{gusynin}
V.~P.~Gusynin, V.~A.~Miransky and A.~V.~Shpagin,
Phys.\ Rev.\ D {\bf 58} (1998) 085023
[arXiv:hep-th/9802136].
%
\bibitem{burden}
C. Burden and A.N. Burkitt, Europhys. Lett. {\bf 3} (1987) 545.
%
\bibitem{SS}
M.~Salmhofer and E.~Seiler,
Commun.\ Math.\ Phys.\  {\bf 139} (1991) 395
[Erratum-ibid.\  {\bf 146} (1992) 637].
%
\bibitem{conformal}
V.~A.~Miransky and K.~Yamawaki,
Phys.\ Rev.\ D {\bf 55} (1997) 5051
[Erratum-ibid.\ D {\bf 56} (1997) 3768]
[arXiv:hep-th/9611142].
%
\bibitem{NJL4}
S.~J.~Hands and J.~B.~Kogut,
Nucl.\ Phys.\ B {\bf 520} (1998) 382
[arXiv:hep-lat/9705015].

\end{thebibliography}
\end{document}